\documentclass[fleqn,10pt]{wlscirep}
\usepackage{amssymb,amsmath,enumerate}
\usepackage{multirow}
\usepackage{bm}
\usepackage{epsfig}
\usepackage{multicol}        
\usepackage{url}
\usepackage{authblk}
\usepackage{bbold}
\usepackage{tabularx}
\usepackage{lscape}
\title{A perspective on correlation-based financial networks and entropy measures}

\author[1]{Vishwas Kukreti}
\author[2*]{Hirdesh K. Pharasi}
\author[3]{Priya Gupta}
\author[4**]{Sunil Kumar}

\affil[1]{School of Computational and Integrative Sciences, Jawaharlal Nehru University, New Delhi-110067, India}
\affil[2]{Instituto de Ciencias F\'{i}sicas, Universidad Nacional Aut\'{o}noma de M\'{e}xico, Cuernavaca-62210, M\'{e}xico}
\affil[3]{Atal Bihari Vajpayee School of Management $\&$ Entrepreneurship, Jawaharlal Nehru University, New Delhi, India}
\affil[4]{Department of Physics, Ramjas College, University of Delhi, New Delhi, India}
\affil[**]{du.sunil@gmail.com, skumar@ramjas.du.ac.in}
\affil[*]{hirdeshpharasi@gmail.com}
\begin{abstract}
\noindent In this brief review, we critically examine the recent work done on correlation-based networks in financial systems. The structure of empirical correlation matrices constructed from the financial market data changes as the individual stock prices fluctuate with time, showing interesting evolutionary patterns, especially during critical events such as market crashes, bubbles, etc. We show that the study of correlation-based networks and their evolution with time is useful for extracting important information of the underlying market dynamics. We, also, present our perspective on the use of recently developed entropy measures such as \textit{structural entropy} and \textit{eigen-entropy} for continuous monitoring of correlation-based networks.
\end{abstract}

\begin{document}

\flushbottom
\maketitle

\thispagestyle{empty}

\section*{INTRODUCTION}
There has been a growing interest in understanding the dynamics of complex systems in the real world. Network science has emerged as an important tool and convenient framework for analyzing a wide variety of social, financial, biological and informative complex systems~\cite{Boccaletti_2006,jackson_2010,barabasi_2016}. Network science began with the seminal papers of Erd{\H{o}}s and R{\'e}nyi~\cite{erdos59a, erdHos_1960}, who proposed random graphs in 1959-60. Random graphs have been used to compare real-world complex networks, since the late 1990s, when a number of scientists started using networks in physical, social, and biological domains. Watts and Strogatz~\cite{watts_1998} renewed the modeling of networks with ``small world'' properties -- random graphs with small diameter but highly clustered like regular lattices. Barabási and Albert investigated the properties of  vertex connectivity of large networks with ``scale-free'' power-law distributions~\cite{barabasi_1999}. These were followed by a flood of papers (see, e.g., ~\cite{barrat2008,caldarelli2007scale,newman2010,albert2002,newman2003,boccaletti2006,lancichinetti2012, gao2016,bian2018,harinder2018}). 
Thus, network science emerged as an important tool for studying different phenomena -- spread of infectious diseases~\cite{Brockmann_2013}, economic development~\cite{Hidalgo_2009, tacchella_2012}, detection, characterisation, identification of long-term precursors to financial crashes~\cite{munnix2012, battiston_2012,Pharasi_2018}, construction of robust sustainable infrastructure and technological networks~\cite{reis_2014}, etc.

Here, we briefly review the role of network science in understanding complex financial markets. Firstly, for uncovering the structure of complex interactions among stocks at a particular instant of time (static picture). For this purpose, one starts with the cross-correlations among stocks returns and then uses various methods of network analysis, such as threshold networks, Minimum Spanning Tree (MST)~\cite{mantegna99information, mantegna1999hierarchical}, Planar Maximally Filtered Graph (PMFG)~\cite{tumminello2005}, etc. Using these methods, one can identify stocks (or sectors) that are strongly or weakly correlated and also study their hierarchy in the network structures. Correlations among stocks change with time, and the underlying dynamics of the market becomes very intriguing. Secondly, a continuous monitoring of financial market becomes very useful and necessary~\cite{Battiston18}, since there are sizable fluctuations during crashes and bubbles. Thus, we discuss here the role of entropy measures in continuous monitoring of the financial market (dynamic picture).  

\section*{CORRELATION-BASED NETWORKS}
  

Mantegna studied the hierarchical structures of correlation-based networks in financial markets \cite{mantegna99information, mantegna1999hierarchical}. Later similar studies of correlation-based networks were made (see, e.g.,~\cite{onnela2002dynamic,Onnela_2003,Bonanno2003topology,shanker_2007a,shanker_2007b}). These correlation-based networks provide easy visual representation of multivariate time series and extract meaningful information about the complex market dynamics. The analysis of evolution of correlation-based networks provides a deep understanding of the underlying market trends, especially during periods of crisis ~\cite{kumar2012correlation,nobi2013network}. We briefly discuss a few methods to construct correlation-based networks from empirical correlation matrix (ECM): MST, threshold network and PMFG.


\subsection*{Minimum Spanning Tree}\label{distancematrix}
MST is constructed by using the distances $d_{ij}=\sqrt{2(1-C_{ij})}$ \cite{Mantegna2003topology,Bonanno2004network,Mantegna2010correlation}, where $C_{ij}$s are the elements of ECM (correlations between pairs of stocks $i,j=1, \dots , N$ in a market for a specific time window), such that all $N$ vertices (stocks) are connected with exactly $N-1$ edges under the constraint that total distance is minimum. Algorithms of Kruskal and Prim are generally utilized to obtain MST from a distance matrix. For a non-degenerate distance matrix, the MST is uniquely determined. Two of the main advantages of MST are that: (i) it produces a network structure without putting any arbitrary threshold, and (ii) it has property of inherent hierarchical clustering. There have been many papers with applications of MST in equity markets~\cite{kumar2012correlation,Sharma2017financial}, currency exchange rates~\cite{jang2011currency}, global foreign exchange dynamics~ \cite{mcdonald2005detecting}. 
Among disadvantages, there is the fact that the order and classification of nodes in a cluster of MST is not robust, and often sensitive to minor changes in correlations or spurious correlations. Therefore, for improvement of results, either noise suppression techniques like Random Matrix Theory (RMT) \cite{plerou2000random} and power mapping \cite{Pharasi_2018} have been used,  or alternative algorithms such as PMFG, Triangulated Maximally Filtered Graph (TMFG), Average Linkage Minimum Spanning Tree (ALMST), Directed Bubble Hierarchical Tree (DBHT) \cite{aste2005complex, tumminello2005tool,song2011nested, song2012hierarchical, tumminello2007correlation} have been proposed. Instead of using pair-wise Pearson correlations, partial correlations and mutual information have also been computed for some studies~\cite{kenett2010dominating, kenett2010dynamics}.

\begin{figure}[t!]
\begin{center}
\includegraphics[width=0.324\linewidth]{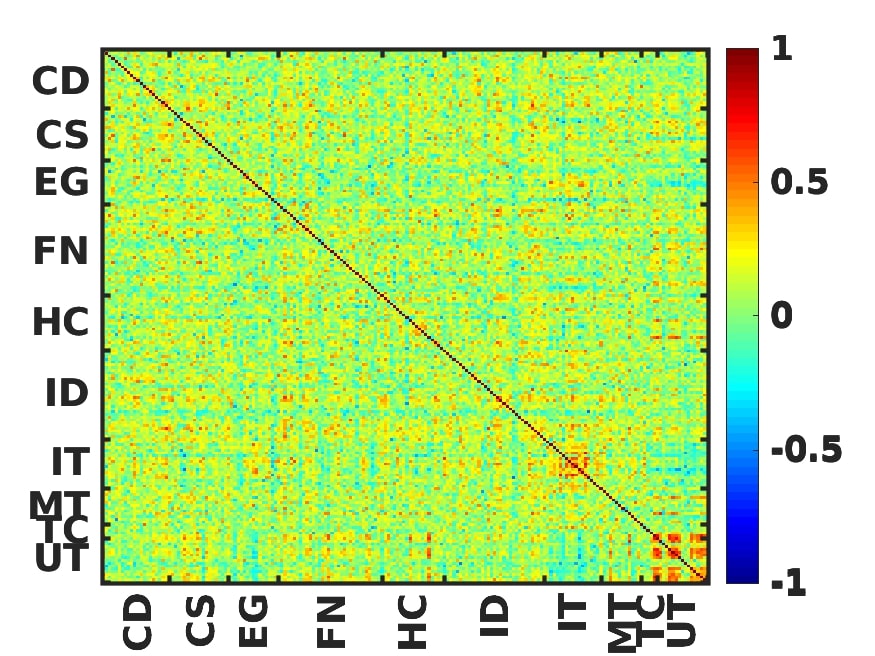}\llap{\parbox[b]{2.2in}{\textbf{A}\\\rule{0ex}{1.7in}}}
\includegraphics[width=0.32\linewidth]{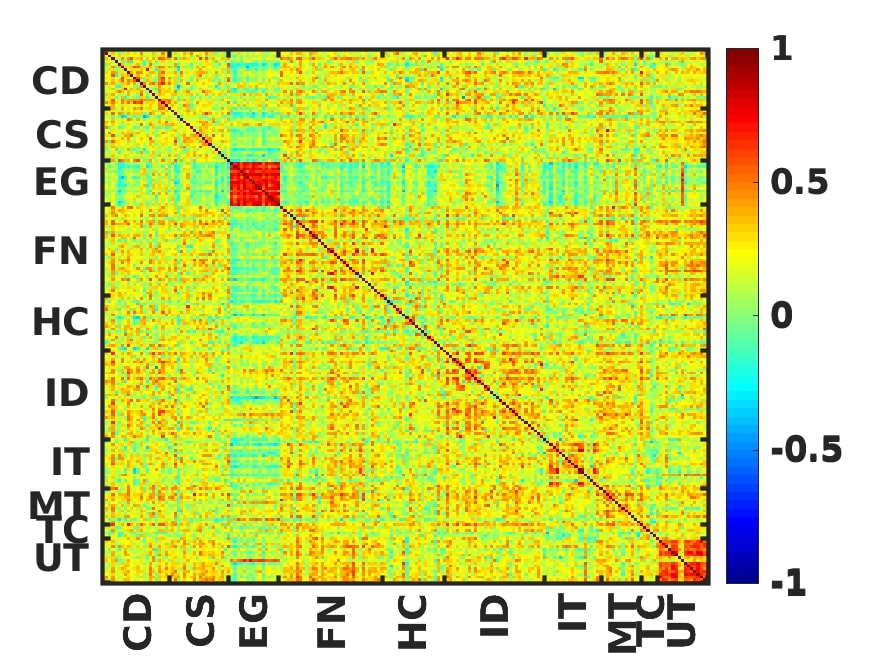}\llap{\parbox[b]{2.2in}{\textbf{B}\\\rule{0ex}{1.7in}}}
\includegraphics[width=0.32\linewidth]{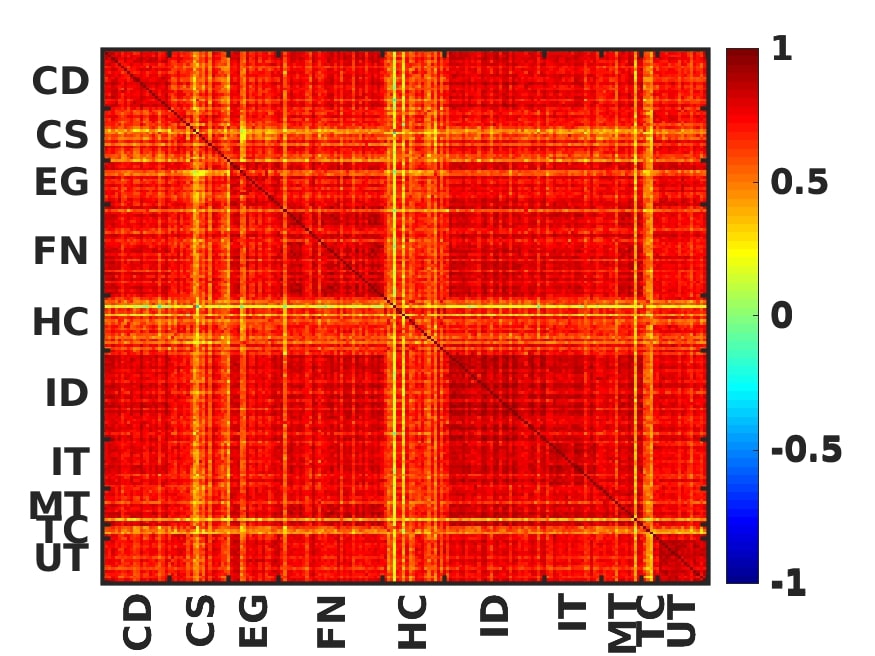}\llap{\parbox[b]{2.2in}{\textbf{C}\\\rule{0ex}{1.7in}}}\\
\includegraphics[width=0.32\linewidth]{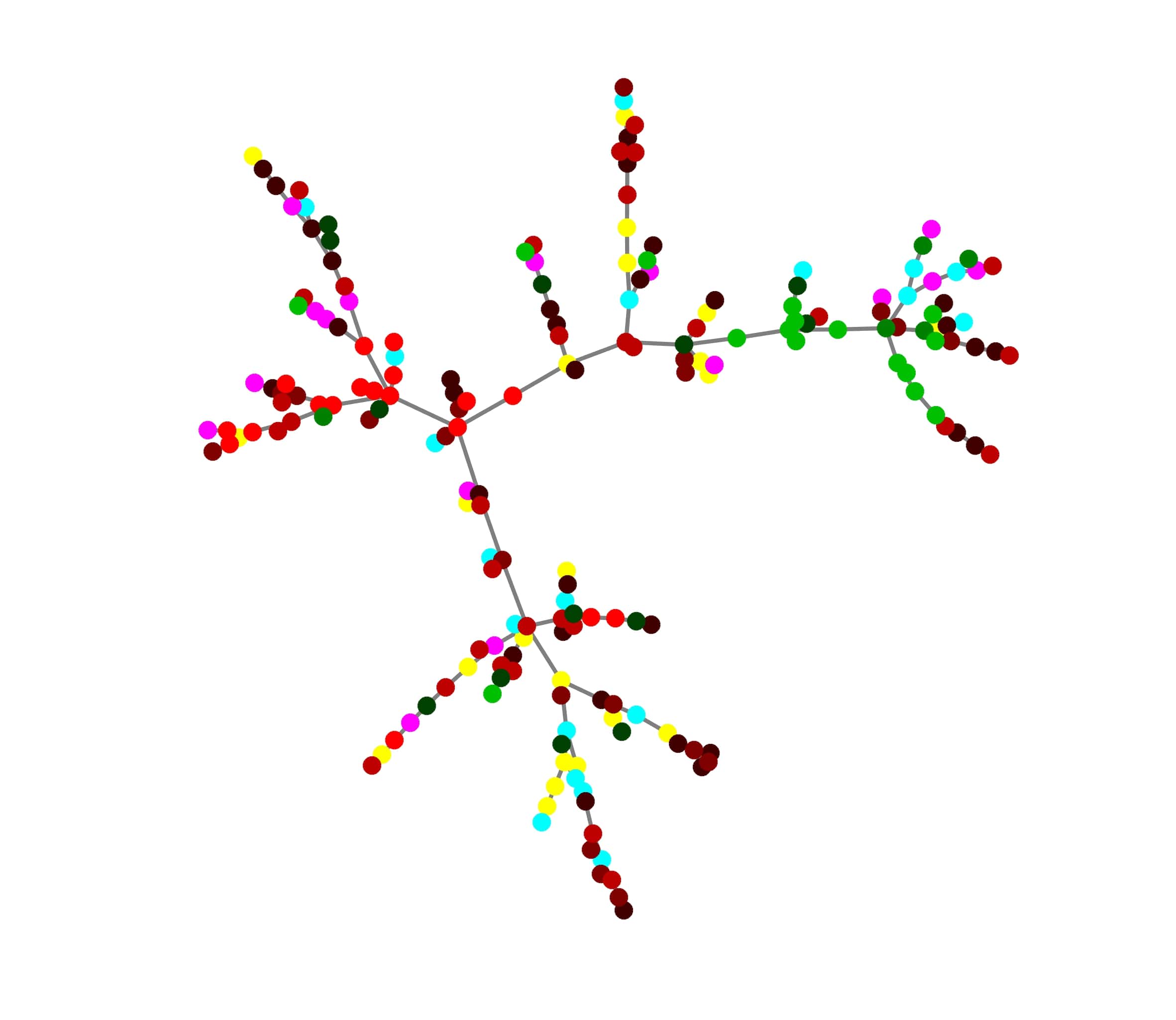}\llap{\parbox[b]{2.2in}{\textbf{D}\\\rule{0ex}{1.7in}}}
\includegraphics[width=0.32\linewidth]{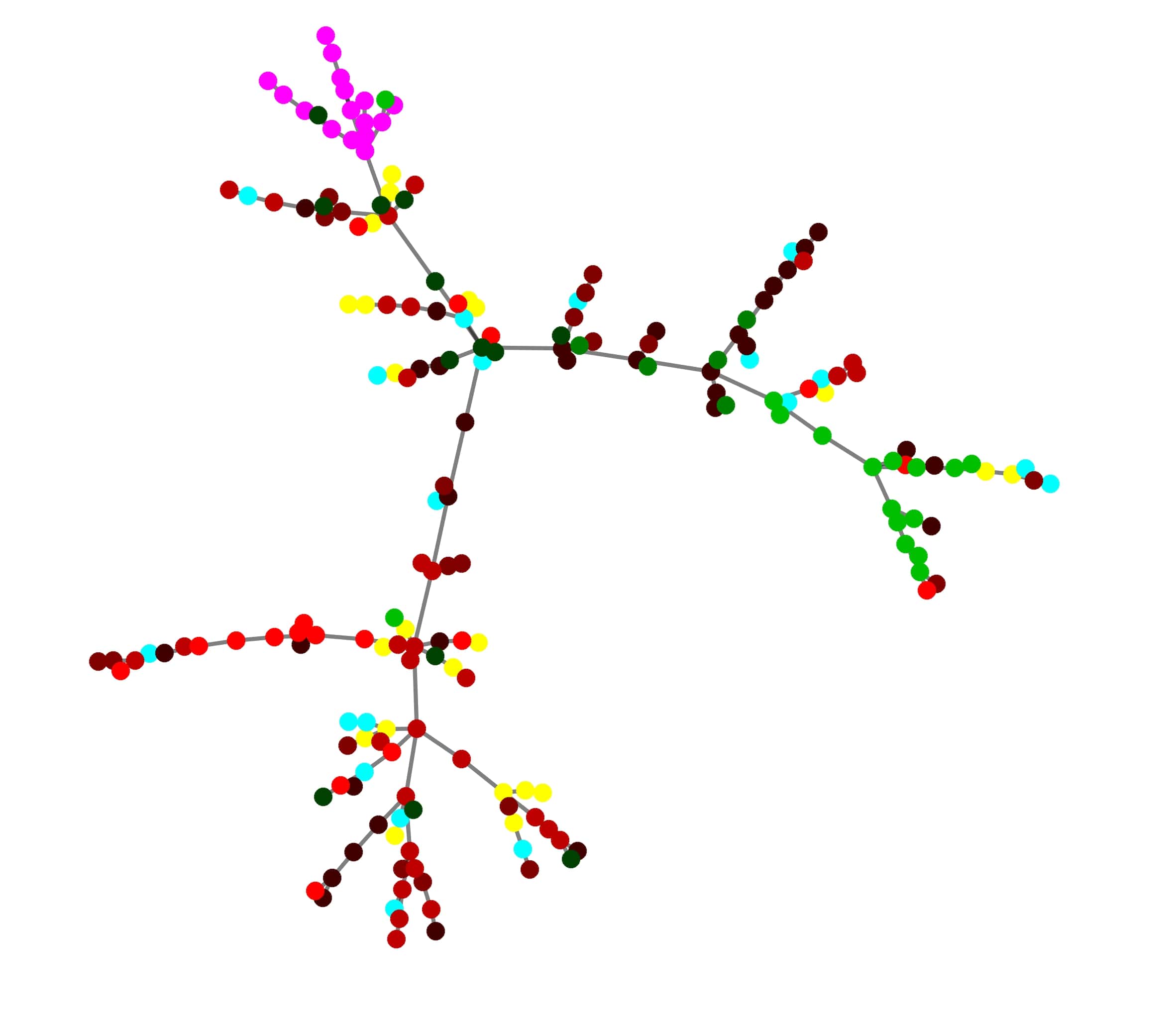}\llap{\parbox[b]{2.2in}{\textbf{E}\\\rule{0ex}{1.7in}}}
\includegraphics[width=0.32\linewidth]{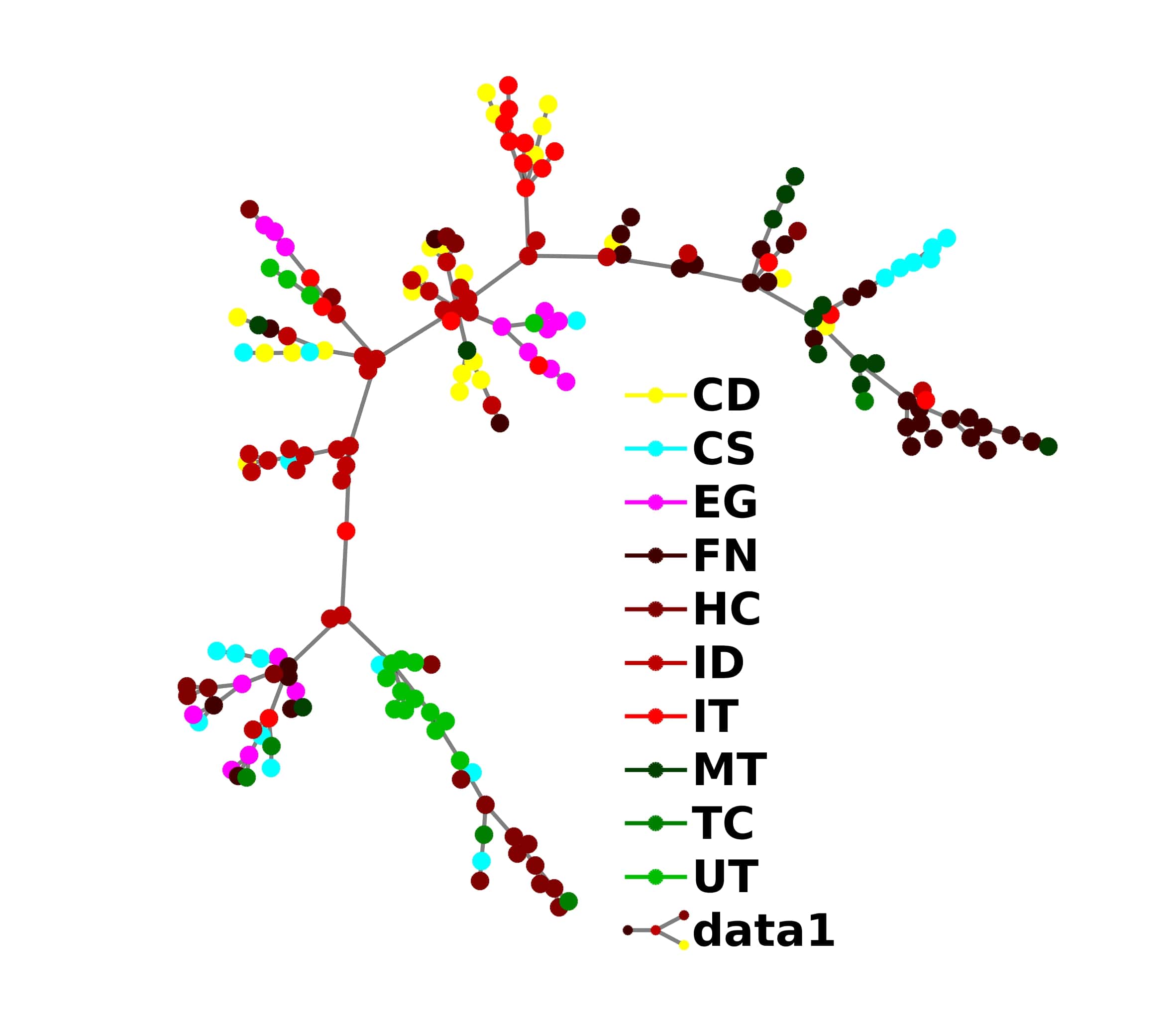}\llap{\parbox[b]{2.2in}{\textbf{F}\\\rule{0ex}{1.7in}}}\\
\includegraphics[width=0.32\linewidth]{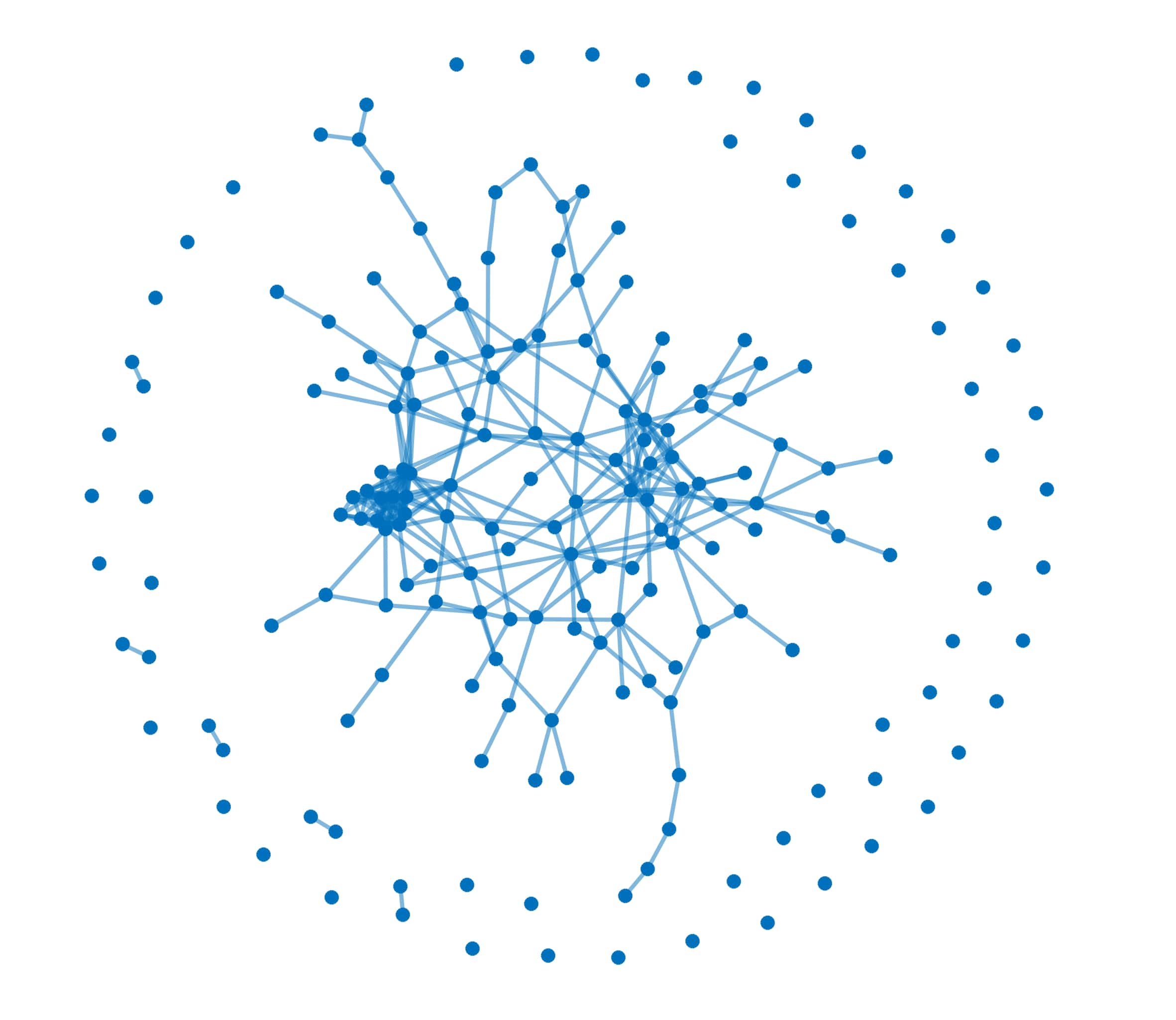}\llap{\parbox[b]{2.2in}{\textbf{G}\\\rule{0ex}{1.7in}}}
\includegraphics[width=0.32\linewidth]{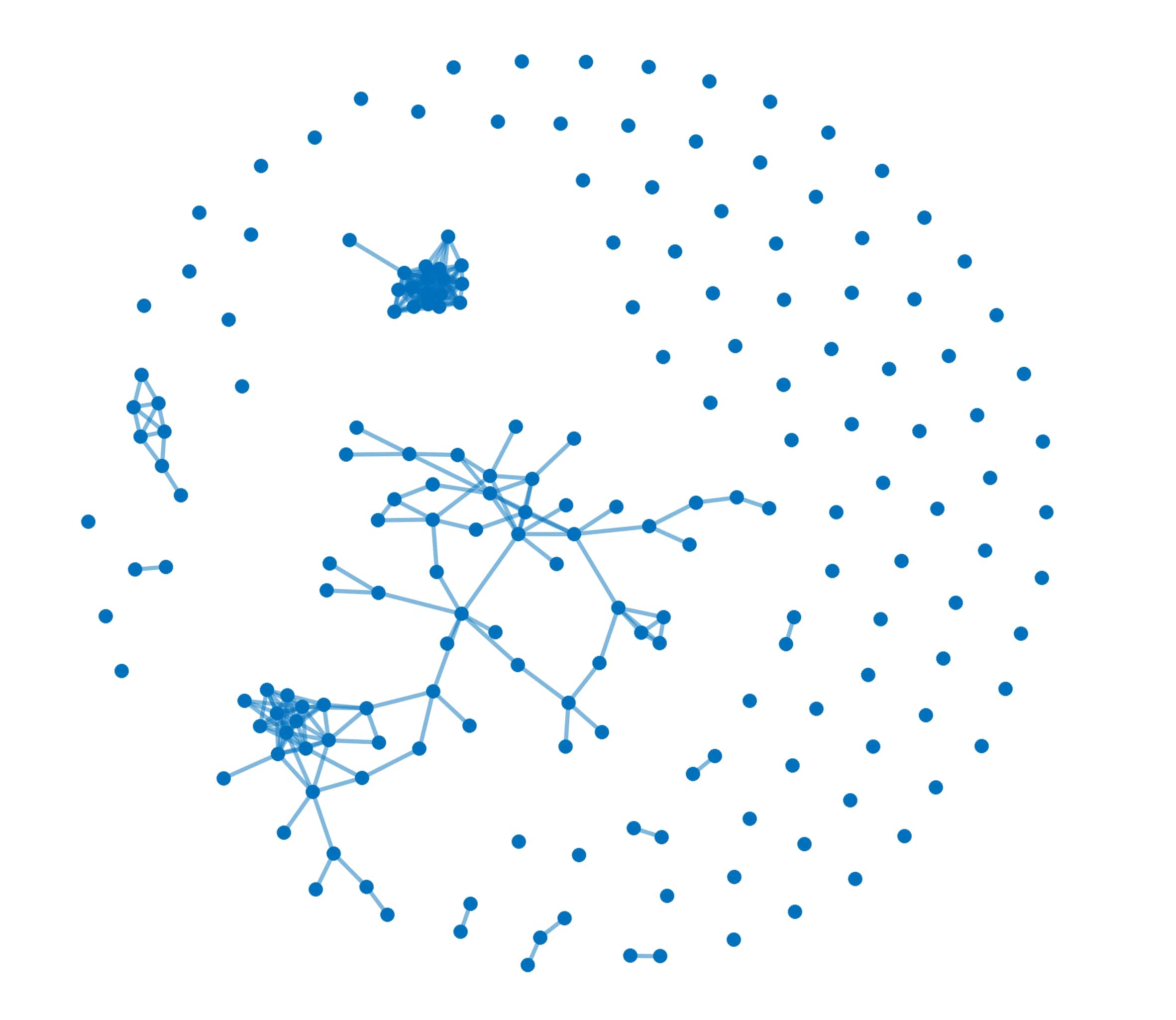}\llap{\parbox[b]{2.2in}{\textbf{H}\\\rule{0ex}{1.7in}}}
\includegraphics[width=0.32\linewidth]{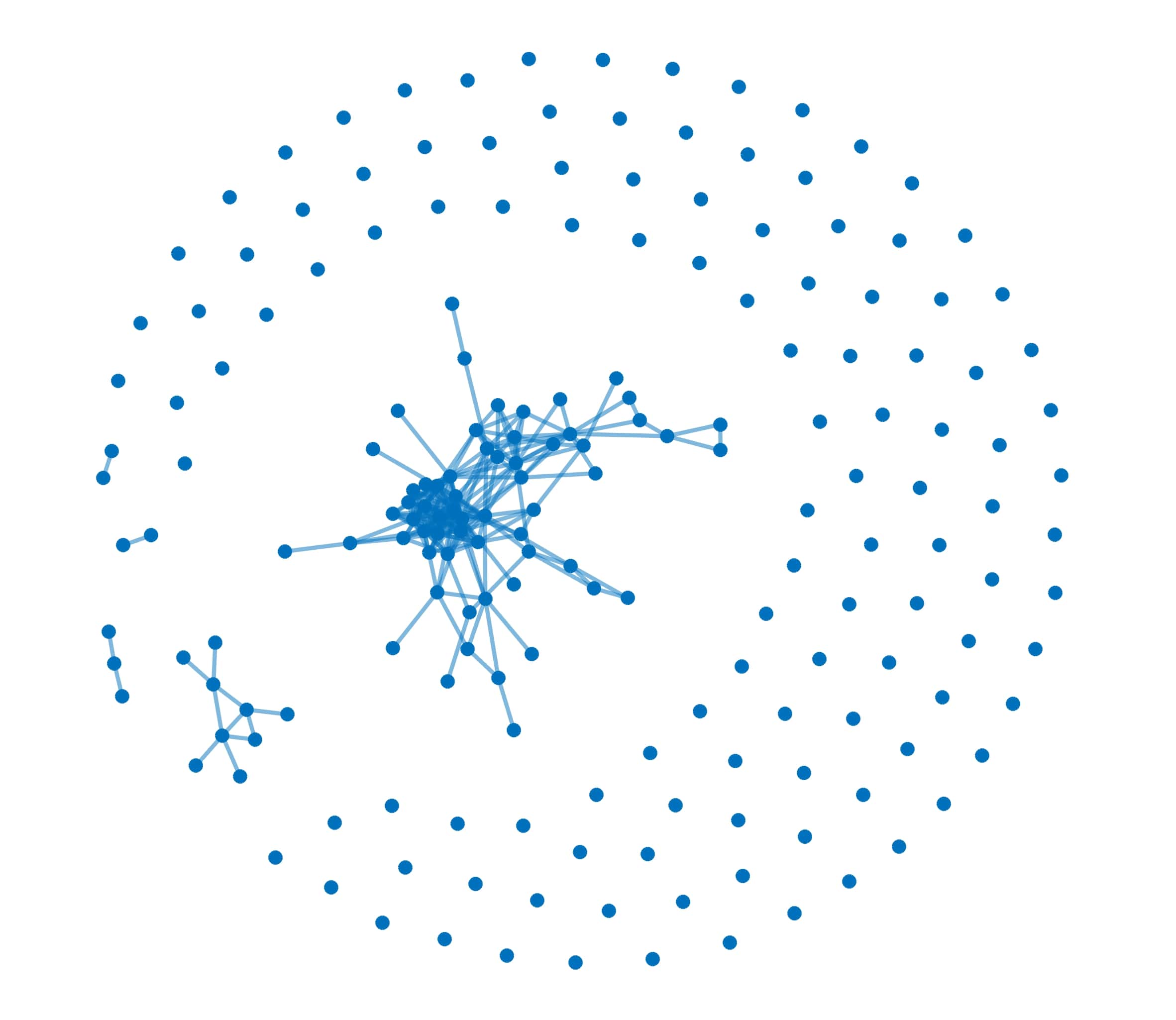}\llap{\parbox[b]{2.2in}{\textbf{I}\\\rule{0ex}{1.7in}}}\\
\includegraphics[width=0.32\linewidth]{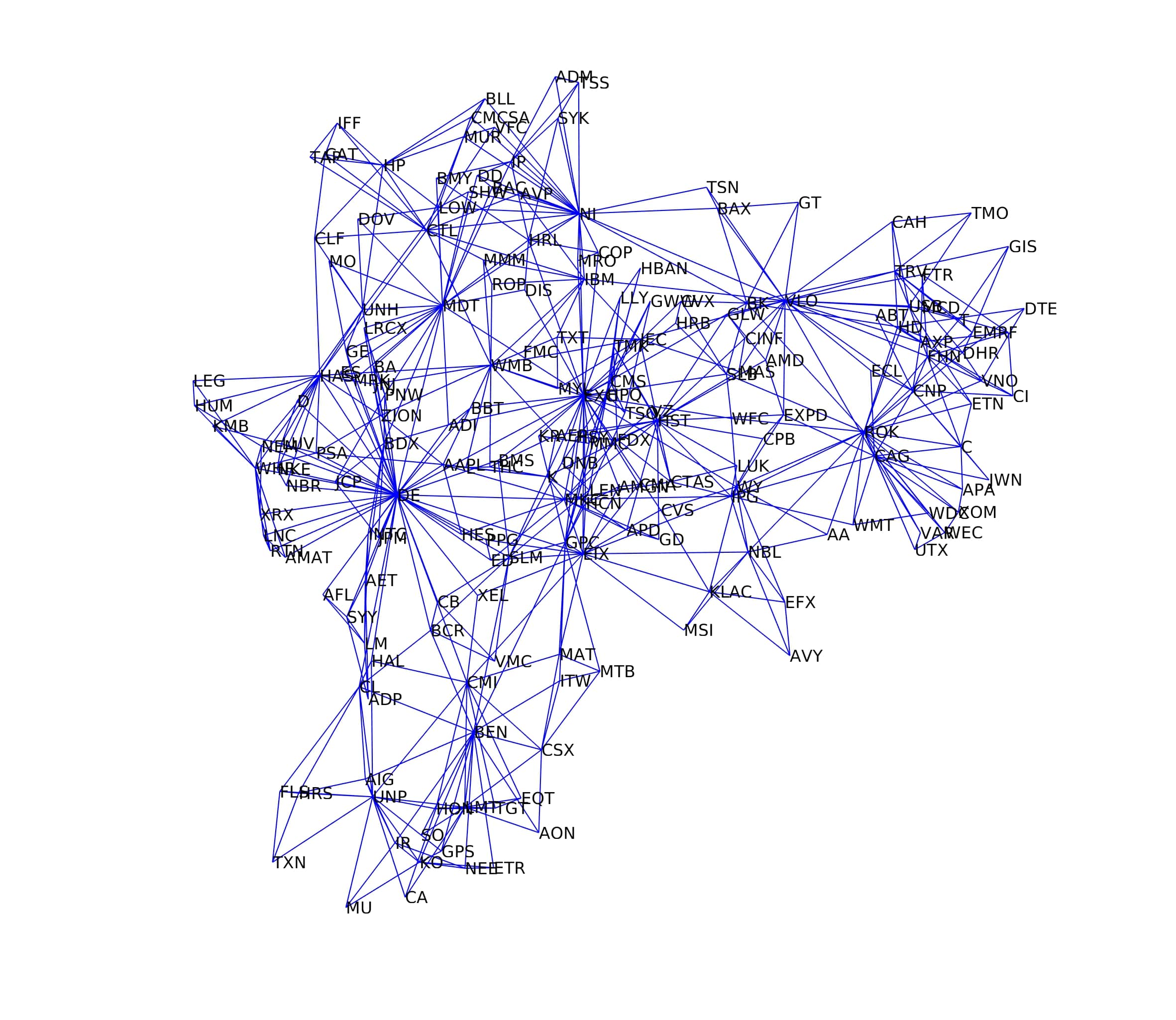}\llap{\parbox[b]{2.2in}{\textbf{J}\\\rule{0ex}{1.7in}}}
\includegraphics[width=0.32\linewidth]{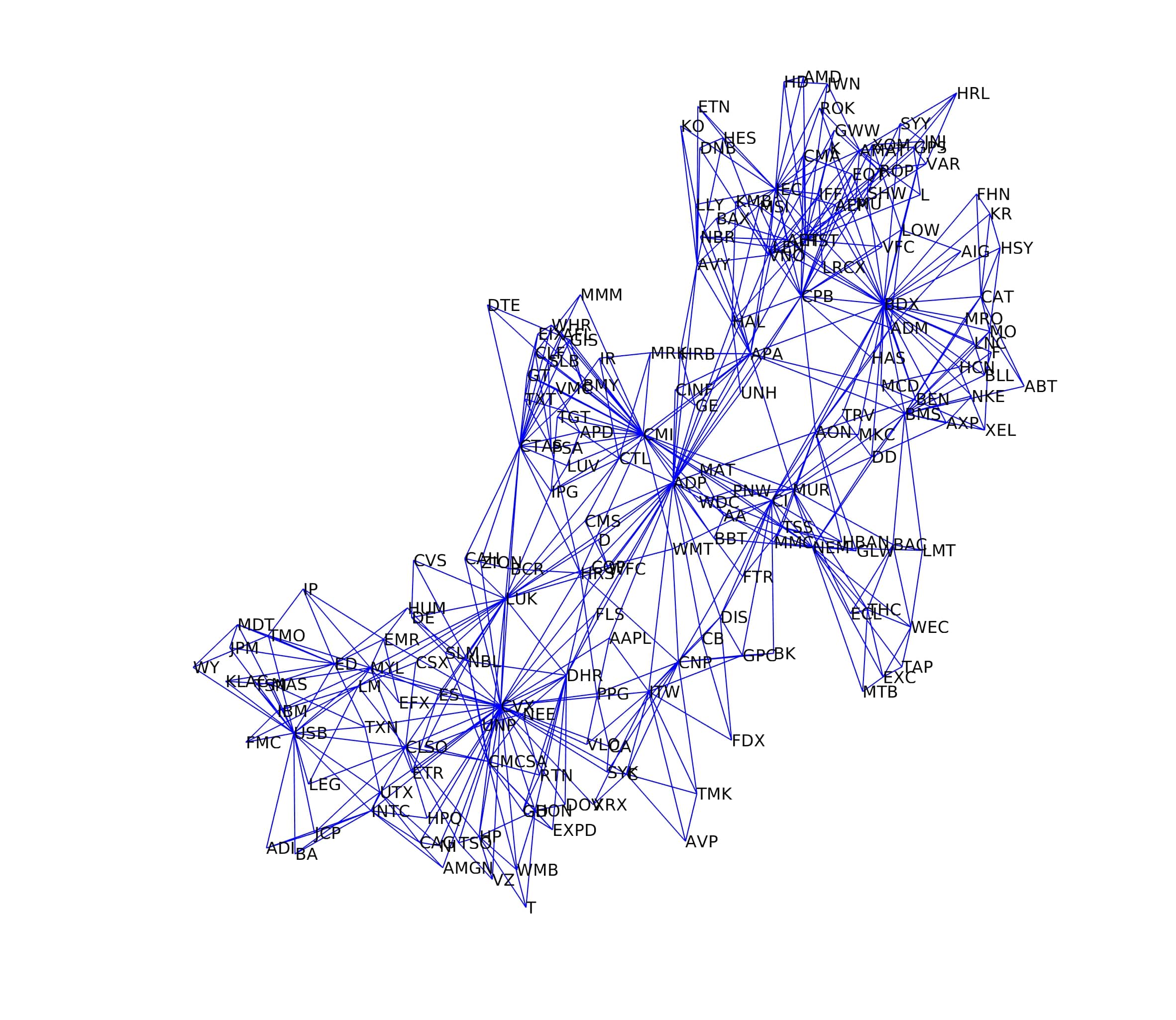}\llap{\parbox[b]{2.2in}{\textbf{K}\\\rule{0ex}{1.7in}}}
\includegraphics[width=0.32\linewidth]{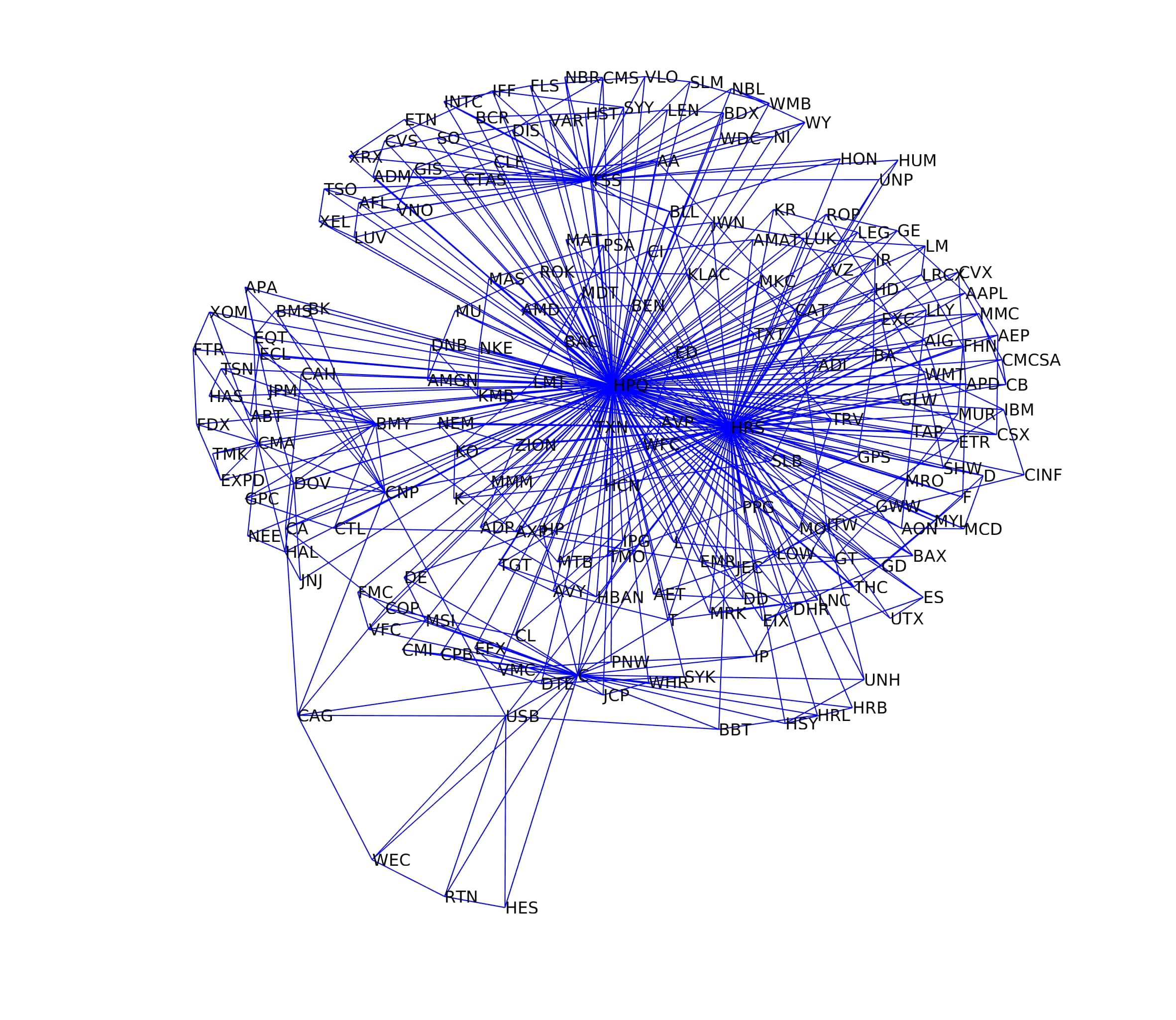}\llap{\parbox[b]{2.2in}{\textbf{L}\\\rule{0ex}{1.7in}}}
\end{center}
\caption{Static correlation-based networks: Analysis of S\&P 500 market with $194$ stocks (epoch of $40$ days) for three different periods: first, second, and third columns are corresponding to 23/07/1985 (normal period), 08/01/2007 (bubble period), and 17/06/2010 (crash period) respectively. \textbf{(A-C)} are heat maps of correlation matrices of different periods. Minimum spanning trees using Prim's algorithm are shown in \textbf{(D-F)}. From \textbf{(G-I)}, correlation based threshold networks at a particular value of threshold. Planar maximally filtered graphs \textbf{(J-L)} for three different periods.There are significant changes in structure of networks in different periods of analysis.}\label{fig:corr-networks}
\end{figure}

MST is useful for studying the taxonomy or the sector classification~\cite{Onnela_2004}, with potential applications in portfolio optimization. 
Researchers have also carried out analysis of dynamical correlations using MST ~\cite{onnela2003dynamics}. This type of dynamical studies has the potential of catching important changes and continuous monitoring of the market. By calculating correlation using rolling window of different lengths, one could construct and analyze the temporal networks. From such analyses, it has been found that configuration of MST structure changes during crisis and there exist strong correlations between normalized tree length and the investment diversification potential \cite{onnela2003dynamic}. 

Figure~\ref{fig:corr-networks} shows the analysis for three periods ending at: (first column) 23/07/1985, (second column) 08/01/2007, and (third column) 17/06/2010. \textbf{Figure~\ref{fig:corr-networks}A-C} show the heat-map of correlation matrices in three different periods, where we have analyzed the S\&P 500 market (consisting of $194$ stocks) with an epoch of $40$ days which is shifted by $20$ days from $1985-2016$. The corresponding MSTs are shown in \textbf{Figure~\ref{fig:corr-networks}D-F}, which have been generated using the Prim's algorithm. Different colors in MSTs correspond to different sectors in the market. One can easily view the changes in the structures of MSTs in different periods of market evolution. 

\subsection*{Threshold Networks}
In this approach, an adjacency matrix is constructed by applying a threshold value in the correlation or fixing the number of edges of the network~\cite{yang2008}. It filters out the strongest correlations by putting a certain value of threshold and discard all remaining correlations below the value of this threshold. A small threshold value gives rise to a completely connected graph, while increasing value of threshold makes the connections less. Thus, one can tune the threshold in order to get the weakly or strongly connected nodes. For a particular value of threshold, as correlation matrices change with time, the threshold networks also change, as shown in \textbf{Figure~\ref{fig:corr-networks}{G-I}} corresponding to the ECMs shown in \textbf{Figure~\ref{fig:corr-networks}{A-C}}. Here the Fruchterman- Reingold (forced-based) layout~\cite{fruchterman1991graph} has been used to visualize the threshold networks. 

One drawback of the threshold networks is that there is a loss of information; when we put a threshold value to the correlation matrix we discard some nodes and edges. Also, threshold networks are very sensitive to the noise (random fluctuations).  However, threshold networks have been constructed and applied in different areas of finance~\cite{wang2008}.  

\subsection*{Planar Maximally Filtered Graph}
PMFG is a network drawn in a plane, such that there are no intersecting links \cite{tumminello2005,nie2018}. If $N$ is total number of stocks, then it contains $3(N-2)$ links. The PMFG has the advantage that it retains the structure of MST (which contains $N-1$ links) and provides additional information about the connections \cite{tumminello2005tool,aste2005complex}. However, PMFG has a disadvantage that there exists a certain arbitrariness in its results, as there is an embedding of data from higher dimension to lower dimension with a zero genus. 
\textbf{Figure~\ref{fig:corr-networks}{J-L}} show the planar maximally filtered graph of matrices shown in \textbf{Figure~\ref{fig:corr-networks}{A-C}}. We find significant changes in the structures of PMFGs in different periods of analysis.

Recently, PMFG and threshold network have been combined to produce PMFG-based threshold networks \cite{nie2017dynamics}. Threshold networks of the financial market are constructed over multi-scale and at multi-threshold~\cite{sui2018}. 

\subsection*{Robustness: Noise suppression and community detection}

We have seen that many of the correlation-based networks have shown clustering with communities of stocks. Thus, community detection in network science serves as an important technique for extraction of the clustering information from ECM of a multivariate time series. Several community detection algorithms have been proposed \cite{macmohan2015community,laloux1999noise,fortunato2010community}.  The problem is that different community detection algorithms yield different results for the same ECM. So, often domain knowledge is required to determine what is a sensible or meaningful community.

Further, we have seen that many of the networks are sensitive to noise or spurious correlations. Properties of random matrices ~\cite{Mehta_2004} have turned out to be useful in reducing noise and thus understanding dynamics of complex systems~\cite{pharasi2019}. An ensemble of random matrices, also known as stationary or standard random (Gaussian) matrix ensemble~\cite{Mehta_2004}, introduced by  Wigner~\cite{wigner_1958,wigner_1967}, have been applied to many studies in physics, biology, finance, etc. (see Refs. \cite{Shukla_2012,Guhr_1998,Alhassid_2000,Kota_2001} and references therein). 
The probability distribution of eigenvalues of Wishart orthogonal ensemble (WOE) follows Marcenko-Pastur distribution \cite{Marcenko_1967}. The ECM of a complex system is normally compared with WOE ~\cite{Plerou2000,Onnela_2003,kumar2015analysing}. It has been observed from eigenvalues statistics of empirical correlation matrices that the few largest eigenvalues show deviations from the Wishart ensemble. Note that 
Pearson cross-correlation  assumes that the time series are stationary, which are valid for shorter lengths of time series. However, if the number of time series are greater than the lengths of time series, then corresponding ECMs are noisy and highly singular. For such short time series, there is a great need of noise suppression in correlation matrix to extract actual correlations. There are different techniques for suppressing the noise in correlation matrix~\cite{guhr2003,bouchaud2000,vinayakpre2013,vinayak_2014}. 
Notably, any ECM of financial market can be decomposed into partial correlations, consisting of market $C_M$, group $C_G$ and random $C_R$ modes, respectively ~\cite{Pan_2007,Sharma2017}. It enables us to identify the dominant stocks, sectors and inherent structures of the market. Recently, detailed analyses of ECMs using these approaches have been carried out to understand the complexity in dynamics of stock market~\cite{Pharasi_2018,pharasi2019,pharasi2020}. It has been found that during the crisis,  the eigenvalue spectrum behaves very differently from one corresponding to a normal period.

\section*{ENTROPY MEASURES}
Entropy measures provide an easy way for continuous monitoring of the financial market, and also prove useful in various other applications in finance, as summarized below. 

Phillippatos and Wilson had used entropy in selection of possible efficient portfolios by applying a mean-entropy approach on a randomly selected 50 securities over 14 years~\cite{philippatos1972entropy}. Using a hybrid entropy model, Xu \textit{et al.} have evaluated the asset risk due to the randomness of the system~\cite{xu2011portfolio}. In 1996, Buchen and Kelly used the principle of maximum entropy for option pricing to estimate the distribution~\cite{buchen1996}, which fitted accurately with a known probability density function. The principle of the minimum cross-entropy principle (MCEP) has been very useful in finance, which was developed by Kullback and Leibler~\cite{kullback1951}. Later, Frittelli discovered sufficient conditions to give a interpretation of the minimal entropy martingale measure~\cite{frittelli2000}. 

Entropy has also been used to understand the financial hazards as well as to construct an early warning indicator for predicting systematic risks~\cite{bowden_2011,gradojevic_2017}. Maasoumi and Racine examined the predictability of the market returns using entropy measure and found that it is capable to detect the nonlinear dependence within the time series of market returns as well as between returns and other prediction variables obtained from other models~\cite{maasoumi_2002}. Recently, Ricci curvature and  entropy have been used to construct an economic indicator for market fragility and systemic risk~\cite{Sandhu_2016}. Very recently, Almog \textit{et al.} presented a perspective on the use of entropy measures such as \textit{structural entropy}~\cite{almog2019}, which is computed from the communities in correlation-based networks. Chakraborti \textit{et al.} computed the \textit{eigen-entropy} from the eigen-vector centrality of the stocks in the correlation-based network~\cite{chakraborti2019}.  
Below, we compare the structural entropy~\cite{almog2019} and eigen-entropy~\cite{chakraborti2019}.

\subsection*{Structural Entropy}
Recently, the concept of Structural Entropy (SE) has been used in monitoring the dynamical correlation based networks of financial market \cite{almog2019}. The SE resolved the problem of choosing different period of crisis and extracting substantial information from the large network of stock market. The SE measures the amount of heterogeneity of the network nodes with an assumption that more connected nodes share common attributes than others. The authors assume the nature of clusters as independent sub-units of the network. The process of calculating the structural entropy involves two steps: (i) Calculation of an optimal partition function which places every node in a certain cluster using a community detection algorithm. (ii) Analysing the partition function and extracting the representative value of the diversity level. Consider a network $G$ with $N$ nodes. The community detection algorithm partitions $G$ nodes in $M$ communities. Let $\sigma$ denote the $N$-dimensional vector where the $i$-th component denotes the community assigned to node $i$. Calculate $M$-dimensional probability vector $P \equiv\left[\frac{c_{1}}{N}, \frac{c_{2}}{N}, \dots, \frac{c_{M}}{N}\right]$, where $c_i$ is the size of community $i$. It is proportional size of the cluster in the network. Then, the formula for Shannon’s entropy is $S \equiv H(P) \equiv-\sum_{i=1}^{M} P_{i} \ln \left(P_{i}\right)$ in terms of probability vector $P$. Structural entropy $S$ of the network provides a way to continuously monitor the state of the network. However, it is sensitive to the choice of community detection algorithm employed in detecting communities. This arbitrariness makes the calculation of entropy dependent on the choice of the user and hence is not universal. 
\subsection*{Eigen-entropy}
Very recently, the concept of eigen-entropy was used in studying financial markets\cite{chakraborti2019}. It is computed from eigenvector centrality of the network obtained from the short time series correlation matrices~\cite{fan2017lifespan, chakraborti2019}. In order to capture the global feature of the network, every node is ranked by its eigenvector centrality and then entropy formula from information theory is used to compute eigen-entropy. Let graph $G(E,V)$ consisting of vertices $V$ and edges $E$. Let $A$ be the adjacency matrix for $G(E,V)$ with $a_{ij}= 1$, if edge connection to vertices $i$ and $j$ are present and $a_{ij}= 0$, if they are not. The sum of all the centralities connected to the vertices is proportional to centrality of the vertex. The adjacency matrix $A$ satisfies the matrix equation $ \mathbf {Ax} =\lambda \mathbf {x}$, where $\lambda$ is the largest eigenvalue of $A$. $A$ is a symmetric positive semi-definite matrix with all non-negative eigenvalues and orthogonal eigenvectors. According to the Perron-Frobenius theorem, any square matrix with all positive entries has a unique solution corresponding to the maximum eigenvalue and its eigenvector with all positive components. Then $v^{\text{th}}$ component of the corresponding eigenvector gives the relative \textit{eigen-centrality} score of the node $v$ in the network. For an absolute score one must normalize the eigenvector, i.e., $ \sum_{i=1}^{N} p_i =1$. The disorderness and randomness of the system uniquely be measured by eigen-entropy and defined as $ H= - \sum_{i=1}^{N} p_i \ln p_i $. Higher the disorder of the system higher the eigen-entropy.
 
Empirical correlation matrix of the market can be decomposed in two logical ways: (i) into three separated modes i.e. market mode $C_M$, the group mode $C_G$ and the random mode $C_R$, where it is arbitrary to chose the range of eigenvalue corresponding to the group mode $C_G$ and the random mode $C_R$ and (ii) into a market mode $C_M$ and group-random modes $C_{GR}$, with no arbitrariness in the system. $C_M ~\& ~C_{GR}$ is the preferable decomposition and corresponding eigen-entropy  $H_M$ and $H_{GR}$ and calculated as $\mathbf {A} = |{C_M}|^2$ (matrix element-wise) and $\mathbf {A} = |{C_{GR}}|^2$ (matrix element-wise), respectively. The eigen-entropy computed using above method gives a simple yet robust measure to quantify the randomness of the financial market without using any arbitrary thresholds. Further Charkraborti \textit{et al.} investigated the relative-entropy which separates the phase space based on their disorder~\cite{chakraborti2019}. The evolution dyanamics of these relative entropies in the phase space show phase-separation with possible order-disorder transitions. These results are certainly of deep significance for the understanding of financial market behavior and designing strategies for risk management.

\begin{figure}[t!]
\begin{center}
\includegraphics[width=0.97\linewidth]{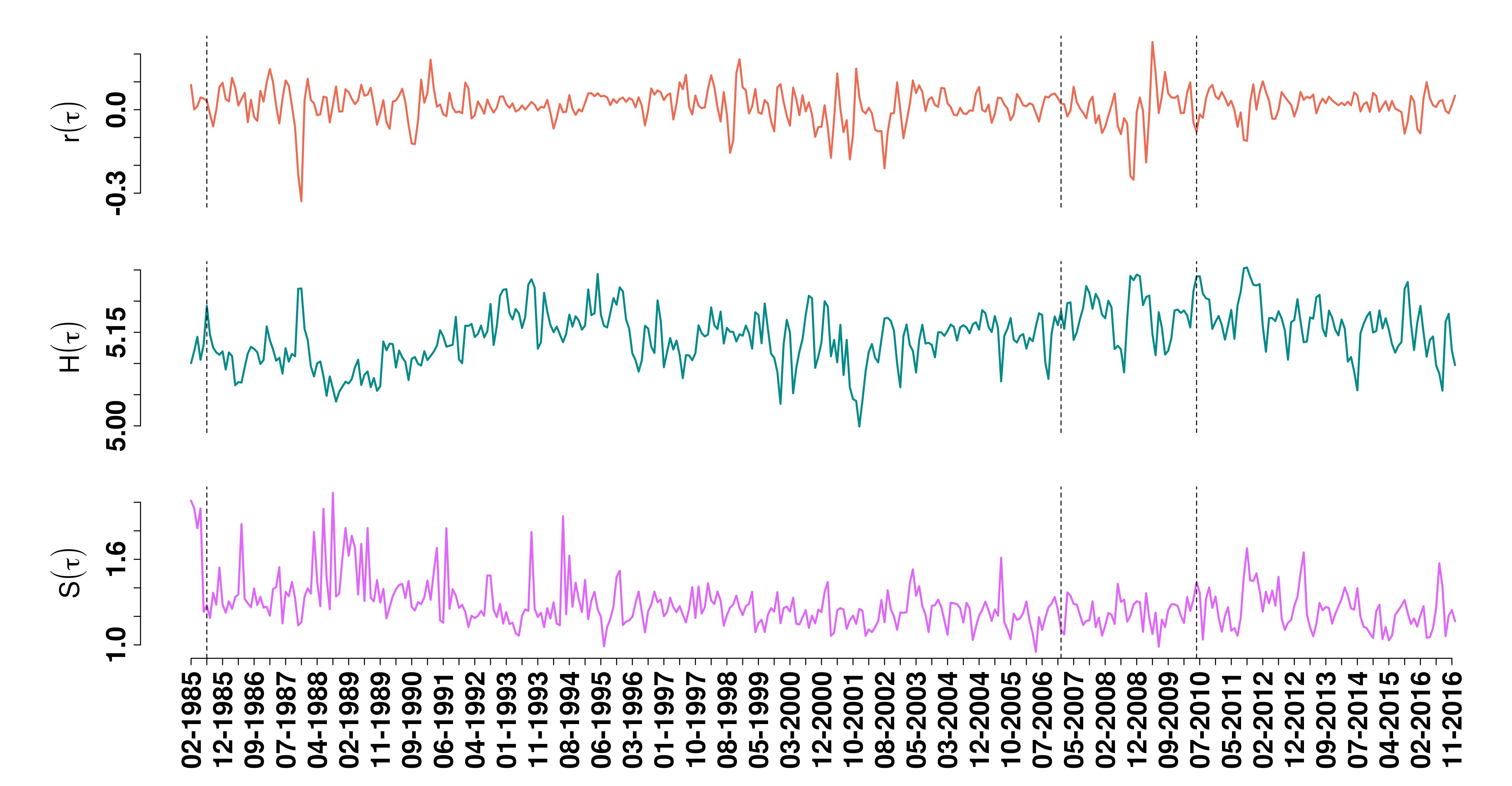}\llap{\parbox[b]{6.1in}{\textbf{A}\\\rule{0ex}{3.2in}}}\llap{\parbox[b]{6.1in}{\textbf{B}\\\rule{0ex}{2.2in}}}\llap{\parbox[b]{6.1in}{\textbf{C}\\\rule{0ex}{1.2in}}}
\end{center}
\caption{Dynamic analysis of stock market (Top to Bottom): Continuous monitoring of S\&P 500 market over a period of $1985-2016$. The logarithmic returns of index is shown in \textbf{A}. Temporal evolution of a new measurement `eigen-entropy' $H(\tau)$, calculated from eigen-vector centralities of filtered correlation matrices (after removal of noise using power map) is shown in \textbf{B}. It can be seen that eigen-entropy easily quantifies the order and disorder in the stock market. Evolution of structural entropy $S(\tau)$ calculated by using community detection algorithm is shown in \textbf{C}. The dashed vertical lines are corresponding to different periods (normal, bubble, and crash) whose static results are shown in Figure \ref{fig:corr-networks}.}\label{fig:timeseries}
\end{figure}
 \textbf{Figure~\ref{fig:timeseries}} shows how the entropy measures can be used for continuous monitoring of the financial markets. \textbf{Figure~\ref{fig:timeseries}A-C} show the evolution of S\&P 500 market over a period of $1985-2016$ for index returns $r(\tau)$, eigen-entropies $H(\tau)$, and structural entropy $S(\tau)$, respectively. Three vertical dashed line are corresponding to epochs ending at 23/07/1985,  08/01/2007, and 17/06/2010.

\section*{CONCLUDING REMARKS}
In this review, we have discussed different methods for analysis of static and dynamic correlation-based networks of financial markets, and also studied how entropy measures can be used to identify normal, bubble, and crash periods. Specifically, we have compared the recently developed concepts of structural entropy and eigen-entropy.

The prediction of collapses of financial markets using traditional economic theories has been a daunting task. These new and alternate methods have the potential use of continuous monitoring and understanding of the complex structures and dynamics of financial markets. These are a few of the attempts physicists have made for generation of early warning signals for crisis, and these methods can be used for timely intervention.

\section*{CONFLICT OF INTEREST STATEMENT}
The authors declare that the research was conducted in the absence of any commercial or financial relationships that could be construed as a potential conflict of interest.

\section*{AUTHOR CONTRIBUTIONS}
SK and HKP designed the idea, wrote the main manuscript text and prepared figures. VK and PG contributed to the literature review. All authors reviewed the manuscript.

\section*{ACKNOWLEDGEMENTS}
The authors are grateful to Anirban Chakraborti, Hrishidev, Suchetana Sadhukhan, Kiran Sharma and Thomas H. Seligman for their critical inputs. 
HKP acknowledges postdoctoral fellowship provided by UNAM-DGAPA. This research was supported in part by the International Centre for Theoretical Sciences (ICTS) during a visit of VK, PG and SK for participating in the Summer research program on Dynamics of Complex Systems (Code:  ICTS/Prog-DCS2019/07).
The topic editors are acknowledged for supporting this open access publication. 

\bibliography{HKP_BIBdoi.bib}

\end{document}